# Matters Arising: On physical nature of the optical de Broglie–Mackinnon wave packets.

Peeter Saari[1,2] and Ioannis M. Besieris[3]

**ARISING FROM** Layton A. Hall and Ayman F. Abouraddy *Nature* Physics https://doi.org/10.1038/s41567-022-01876-6 (2023)

Hall and Abouraddy [1] have reported first experimental observation of optical de Broglie–Mackinnon wave packets, which is a seminal achievement in the study of so-called non-diffracting optical pulses. These wave packets propagate in free space without spreading with subluminal relativistic velocities, i.e., with speeds slower but close to the velocity of light in vacuum. The experiments in [1] became possible thanks to the application of quite a witty method. Unfortunately, the explanation of the physical nature of the wave packets and their graphical and mathematical descriptions in the theoretical part of [1] suffer from some ambiguities that need to be clarified.

De Broglie–Mackinnon wave packets (henceforth shortly 'dBM wave' as in [1]) have been theoretically studied for decades and their physical nature is well understood [2-7]. Essentially, a free-space dBM wave is a monochromatic spherical standing wave observed from another, inertially moving frame. Or *vice versa*— the standing wave is generated in a moving frame and observed in the stationary (laboratory) frame. In both situations detectors see the wave as an invariantly propagating localized wideband pulse. What is important here, is that the dBM wave constitutes a source-free wavefield, i.e., its sources are remote. It can be treated as a superposition of two waves—one incoming toward the centre and another outgoing from the centre. Due to the mutually opposite signs of these waves, there are no near-field features or a singularity at the centre of the wave as is in the case of, say, a dipole-generated field. In short, source-free wavefields can be described as generated by a pair of a source and sink at the same location (see, e.g., [8-9]).

Paper [1] deals with two-dimensional (2+1) dBM waves, i.e., cylindrical waves instead of tree-dimensional ones. Nevertheless, the basic properties mentioned above apply as well. Despite the fact that the intensity of (2+1) dBM waves remains nonlocalized in one (say, *y*) dimension and they do not represent the original dBM wave packet [3], the pioneering contributions in [1] are undeniable. Most likely, Ref. [1] will be widely read (by professionals as well as students) and intensively cited for decades as it has happened with the optical realization of the X wave [10]—a superluminal counterpart of the subluminal dBM wave (Ref. [10] as well as [6] and [7] are referred to in [1]). All the more, the fallacies in [1] need clarification which we undertake in what follows.

    1. Throughout the theoretical part of [1] the relevant fields are explained as if generated by a dipole source at the origin (see, e.g., Fig. 2). For the reasons given above, if this were the case, the field would not look like the plot in the last column of Fig. 2a which depicts a standing *source-free* field as it should be.

    2. The plot in the last column of Fig. 2b—which is very important for understanding the formation of dBM waves—is misleading because it does not depict a dBM wave but something else.

---

[1] Institute of Physics, University of Tartu, W. Ostwaldi 1, Tartu 50411, Estonia.
[2] Estonian Academy of Sciences, Kohtu 6, Tallinn 10130, Estonia.
[3] The Bradley Department of Electrical and Computer Engineering, Virginia Polytechnic Institute and State University, Blacksburg, VA 24060, USA.

The plot shows that wavelengths ahead and behind the pulse are different due to the Doppler effect. In dBM waves there is no such a difference, although the aberration and the Doppler effect play a decisive role in the formation of any dBM waves [6]. In any point of a dBM wave the incoming and outgoing constituents have opposite directions and therefore the Doppler shift does not manifest itself in the way depicted in the plot. The plot would depict the outgoing field—affected by the Doppler effect—of a moving source, had it a peculiarity at the location of the source.

3. A minor discrepancy is between the caption of Fig. 2 and its last column: the former says that the plot depicts "*the real part of the spatio-temporal field profile ψ(x, z; t) at a fixed axial plane z*," while the designations in the plot indicate that the column shows the modulus squared of ψ(x, z; t) at a fixed *time instant*.

4. A principal mistake is in Eq. (2). In the role of the factor representing the subluminally and rigidly propagating envelope of the dBM wave packet, there is the sinc-function (which is also known as the zeroth-order *spherical* Bessel function). It would be correct in the case of (3+1)-dimensional, but not for the (2+1)-dimensional dBM wave packet considered in [1]. Therefore, the wave function given by Eq. (2) does not satisfy the wave equation it must obey. Instead of the sinc-function, Eq. (2) must contain the zeroth-order *cylindrical* Bessel function of the first kind $J_0(.)$ with the same argument expression (without division by π). The function $J_0(.)$ results from a straightforward evaluation of the superposition integral over plane waves with the spectrum considered in [1] and depicted in Fig. 2. It is generally known that the Fourier transform of a spectrum like that considered in [1] is the Bessel function $J_0(.)$ The latter and the sinc-function have qualitatively similar oscillatory behaviour. However, with the distance $r$ from the maximum, the sinc-function decays as $r^{-1}$ whereas $J_0$ decays as $r^{-1/2}$, i.e., substantially slower.

Finally, we point out that the dBM wave packet may move with arbitrary subluminal speeds, not necessarily with relativistic ones. Moreover, any stationary focused monochromatic light field, if observed by an appropriate detecting apparatus which moves arbitrarily slowly with respect to the laboratory table, sees a dBM wave-like pulse when the detector passes through the focus.

In conclusion, while the critical comments do not reduce the merits of the work [1]—especially of its experimental part—, they will hopefully help to comprehend the physics of the optical dBM wave packets.